\documentclass[conference]{IEEEtran}
\usepackage{cite}

\ifCLASSINFOpdf
  \usepackage[pdftex]{graphicx}
\else
\fi
\usepackage{fixltx2e}

\usepackage{stfloats}

\usepackage{lipsum}
\usepackage{xcolor}

\begin{document}
%
\title{AVAC: A Machine Learning based Adaptive RRAM Variability-Aware Controller for Edge Devices}



%
\author{\IEEEauthorblockN{Shikhar Tuli\IEEEauthorrefmark{1} and
Shreshth Tuli\IEEEauthorrefmark{2}}
\IEEEauthorblockA{\IEEEauthorrefmark{1}Department of Electrical Engineering, Indian Institute of Technology Delhi}
\IEEEauthorblockA{\IEEEauthorrefmark{2}Department of Computer Science and Engineering, Indian Institute of Technology Delhi}
\IEEEauthorblockA{Email: \{shikhartuli98, shreshthtuli\}@gmail.com}
}


\renewcommand\IEEEkeywordsname{Keywords}

\maketitle

\begin{abstract}
Recently, the Edge Computing paradigm has gained significant popularity both in industry and academia. Researchers now increasingly target to improve performance and reduce energy consumption of such devices. Some recent efforts focus on using emerging RRAM technologies for improving energy efficiency, thanks to their no leakage property and high integration density. As the complexity and dynamism of applications supported by such devices escalate, it has become difficult to maintain ideal performance by static RRAM controllers. Machine Learning provides a promising solution for this, and hence, this work focuses on extending such controllers to allow dynamic parameter updates. In this work we propose an Adaptive RRAM Variability-Aware Controller, AVAC, which periodically updates Wait Buffer and batch sizes using on-the-fly learning models and gradient ascent. AVAC allows Edge devices to adapt to different applications and their stages, to improve computation performance and reduce energy consumption. Simulations demonstrate that the proposed model can provide up to 29\% increase in performance and 19\% decrease in energy, compared to static controllers, using traces of real-life healthcare applications on a Raspberry-Pi based Edge deployment.
\end{abstract}



%
\IEEEpeerreviewmaketitle

\section{Introduction}

With the rapid technology burst in the Internet of Things (IoT) domain, the demands from Edge and Fog devices have been continuously increasing \cite{shi2016edge}. Edge devices are becoming more complex, expected to do more-and-more computations in order to reduce the data to be sent through the transceiver; thus reducing the bandwidth requirement, latency and energy consumption. With the eventual goal of a completely independent Edge node \cite{Gill2019}, designers are constantly making energy specifications stricter for Edge devices \cite{shi2016edge}. Further, with CMOS technology scaling to the deep sub-micron domain, the leakage power has become comparable to the logic power consumption \cite{itrs2011}. To alleviate this problem, it is lucrative to enhance the performance of such devices and then expect them to quickly switch to the idle deep-sleep mode \cite{Braojos2016}.

In that context, emerging Resistive Random Access Memory (RRAM) technologies serve as a good candidate as opposed to the traditional eFlash technologies, thanks to their non-volatile operation, zero leakage, easy co-integration with CMOS process, low programming voltage and fast switching \cite{vianello, wongRRAM}. RRAMs find applications in novel system designs with non-volatile cache and universal memories \cite{Torres1, Torres2}. These new architectures can allow sufficient gains in performance and energy, much needed for upcoming Edge/Fog devices.

One such architecture is the Static RRAM Variability-Aware Controller (RRAM-VAC) proposed by S. Tuli et al. in \cite{Tuli_aspdac}. To tackle the problem of high device-to-device and cycle-to-cycle temporal variability in RRAMs (which can be up to several decades \cite{Sassine2018}), the Static RRAM-VAC utilizes the recently proposed Write Termination (WT) circuits \cite{AlayanWT}. Thanks to their capability of dynamically detecting and stopping the programming operation once the device has switched, the Static RRAM-VAC can coalesce multiple write operations before triggering the write request. Thus, it can average out the temporal write-time variability and effectively run the system at the memory programming time distribution mean rather than the worst case tail. However, with a fixed and rigid architecture, the Static RRAM-VAC is not robust and adaptive enough to accommodate widely varied and unpredictable memory trace patterns in developing complex Edge/Fog devices. 

Hence, in this work we propose a Machine Learning (ML) based approach to dynamically tune operation parameters so as to maximize performance and limit energy requirement at all times of operation. For this, we characterize different Wireless Body Sensor Node (WBSN) applications, a typical candidate for large amount of critical data which needs to be processed timely, securely and efficiently. We exploit a polynomial regression technique to model the operation parameters, namely the Wait Buffer and batch sizes with the performance and energy gains. We then apply gradient ascent method to maximize the gains dynamically, thus providing up to 94\% performance gains (up to 29\% more than the Static RRAM-VAC) and up to 99\% energy gains (up to 19\% more than the Static RRAM-VAC) on traces of real-life healthcare applications, including the HealthFog \cite{tuli2019healthfog, tuli2019fogbus} framework.


The rest of the paper is organized as follows. Section \ref{sec:bck} presents the Static RRAM-VAC operation and the Machine Learning technique used. Section \ref{sec:arch} shows the proposed AVAC architecture. Section \ref{sec:ExpSetup} describes the RRAM technology assumptions and experimental setup used for simulations. Section \ref{sec:ExpResults} compares the gains of the Adaptive over the Static RRAM-VAC. Finally, Section \ref{sec:Conc} concludes the paper.

\section{Background}
\label{sec:bck}

\subsection{The Static RRAM-VAC}
\label{sec:Static_RRAM-VAC}

The Static RRAM Variability-Aware Controller (RRAM-VAC) proposed in \cite{Tuli_aspdac} stores the write requests and locks them to form a batch. This batch is then written to an RRAM using the ``Write Coalescing" method - as part of a single operation, writing the next bit as-soon-as the previous one is finished \cite{Tuli_aspdac}. This allows it to effectively interact with a synchronous processor and write to the RRAM asynchronously, further enabling the system to run at the mean of the switching-time distribution rather than at the worst-case tail.

Functionally, the Static RRAM-VAC uses a Wait Buffer and a Read Buffer. The Wait Buffer is implemented as a Binary Content Addressable Memory (BCAM) \cite{AgarwalBCAM} of a fixed size. The locked batch is a part of the Wait Buffer and cannot catch read requests. On the other hand, the rest of the Wait Buffer can still catch both read and write requests. Read requests are first sent to the Wait Buffer. If the corresponding address is not present in the Wait Buffer, the request is issued to the RRAM. If a locked batch is in process, the read request is stored in the Read Buffer where it waits until the RRAM is available.

However, the given system has a fixed size of the Wait Buffer and batch, optimally dependent on application patterns \cite{Tuli_aspdac}. With the increasing complexity of Edge/Fog workloads, there is a need to dynamically tune the Wait Buffer and batch sizes in order to enhance energy and performance. Moreover, these workloads correspond to real-life applications and continuously engage with unpredictable service demands and tasks. This makes prediction of optimum buffer and batch sizes difficult. Thus, in this work, we propose an on-the-fly ML based approach of dynamically tuning these sizes in order to maintain the best performance and energy characteristics.

\begin{figure}
    \centering
    \includegraphics[width = 0.9\columnwidth]{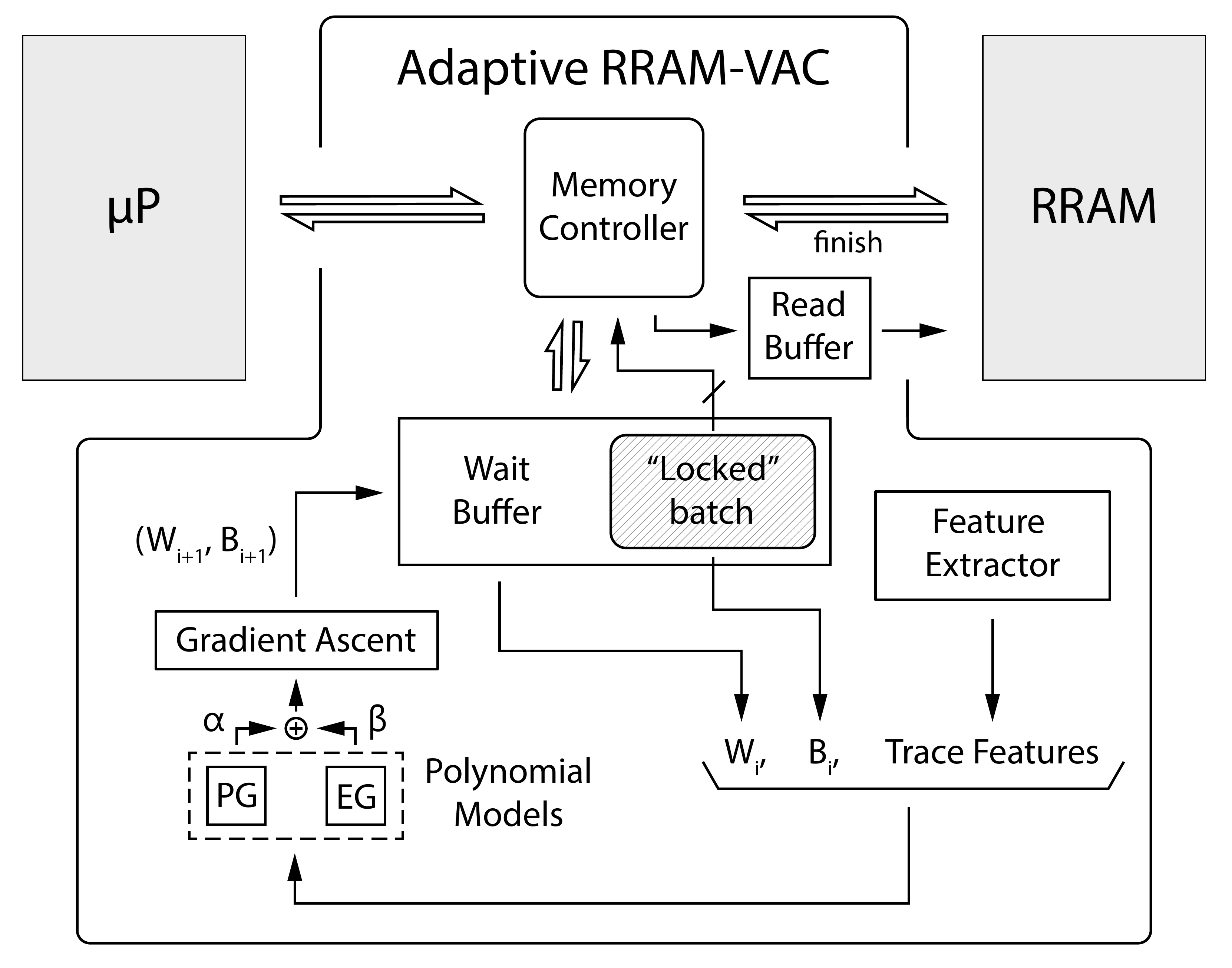}
    \caption{Proposed Adaptive RRAM-VAC (AVAC) Block diagram}
    \label{fig:block_diagram}
\end{figure}

\subsection{Machine Learning models}
\label{sec:ml_models}

The proposed AVAC model uses polynomial regression \cite{montgomery2012introduction} and gradient ascent \cite{bubeck2015convex}. The former provides a way to approximate a function with input as a multidimensional vector, characteristic to the application and the current Wait Buffer and batch sizes. We use this to approximate the expected performance and energy gains and determine a cumulative reward as described in Section \ref{sec:ml}. We then use gradient ascent to iteratively update the Wait Buffer and batch size pair, in order to optimize this reward function.

\section{Adaptive RRAM-VAC Architecture}
\label{sec:arch}

This section gives a functional description of AVAC.

\subsection{Dynamic Buffer Size}
\label{sec:Dynamic_Buffer_Size}

Figure \ref{fig:block_diagram} presents a detailed block diagram of the proposed AVAC. As explained in Section \ref{sec:Static_RRAM-VAC}, the controller fetches the read/write requests from the processor and stores them to the Wait Buffer (for a write request) or the Read Buffer (for a read request). The ``locked" batches are written employing the ``Write Coalescing" technique. However, in this architecture, the Wait Buffer is larger in size and we can dynamically switch off certain Write-Lines based on the requirement. This allows us to change the effective Wait Buffer size that is being used at any given time. The power-gating reduces leakage in applications where a large size is not needed, since it shuts off the current to these cells. On the other hand, for applications with very low data locality, we can optimize performance by increasing the Wait Buffer size, since that would better average the variability in write times.

Further, the whole Wait Buffer can be connected to the RRAM block via a wide bus. We can dynamically tune the batch size by varying the number of words written at a time. Higher variability ($\sigma$) of the distribution requires a larger batch size. Moreover, the optimum batch size also depends on the Wait Buffer size \cite{Tuli_aspdac}.

\subsection{Feature Extraction}
\label{sec:fe}

The optimum Wait Buffer and batch sizes and the possible performance and energy gains are dependent on the memory-trace features, as introduced in Section \ref{sec:Static_RRAM-VAC}. These features represent the application memory access patterns. They are extracted in the AVAC by the ``Feature Extractor" as shown in Figure \ref{fig:block_diagram}. The set of all the 8 features, with the current \textit{Wait Buffer} size ($W$) and \textit{batch} size ($B$) forms a 10 dimensional feature vector. The 8 features extracted are as follows:

\begin{enumerate}
    \item Read/Write ratio: Large number of reads requires a small batch size, so that read requests do not have to wait for a long time for the batch to flush out to the RRAM. If the read locality is high, then a small Wait Buffer can also lead to performance gains (reading from the Wait Buffer is less expensive than from the RRAM block).
    \item Read locality: As explained above, high read locality would benefit from a small Wait Buffer. If the locality is low, the application would demand a larger Wait Buffer.
    \item Write locality: Again, high write locality would benefit from a small Wait Buffer just as the reads above.
    \item Mean Read burst size: For a series of small read bursts, the Wait Buffer will have to wait frequently for the Read Buffer to clear out, thus not allowing it to flush out its write requests so as to catch the next write instruction.
    \item Mean Write burst size: While the batch is being written to memory, the next read request that cannot be processed inside the Wait Buffer is stored in the Read Buffer and the processor is stalled, compromising performance.
    \item Mean Read repetition: The read locality only targets the variance in the read request addresses. The peak of the distribution can be represented by the mean of the repetition for each read address.
    \item Mean Write repetition: Similar to the case of read repetition, higher write repetition would result in higher energy gains by accesses from the Wait Buffer.
    \item Variation in bit-changes for writes: By this feature, we mean the number of changes in bit written to a particular address (0 to 1 or 1 to 0). The Write Coalescing strategy works best if the variability in writes can be averaged out. However, for example, if the writes to addresses are of the form 0x00000000 to 0x0000000F, then the LSB will always have higher write times than the other bits, thus reducing the gains of a higher batch size. 
\end{enumerate}

\subsection{Polynomial model and parameter optimization}
\label{sec:ml}

The complete feature vector, calculated for a fixed number of memory accesses in 10 dimensions, is given to two polynomial-fit based models: Performance Gains (PG) and Energy Gains (EG). We empirically find 1000 to be the best interval size ($S_I$), and propose to test with other and dynamic $S_I$ values as part of future work. Each model is an approximator that maps the 10 dimensional feature vector ($fv$) to the corresponding performance and energy gains ($pg$ and $eg$). At training time, we generate multiple data-points \{$fv^{(j)}$, $pg^{(j)}$, $eg^{(j)}$\}$_{j=1}^N$ for different applications with buffer size $\in[1,120]$ and batch size $\in[1,80]$. This training data is then used to perform polynomial regression and generate model parameters for PG and EG. The applications were chosen to cover a large space of feature vectors.

\begin{figure}[!t]
    \centering
    \includegraphics[width = 0.9\columnwidth]{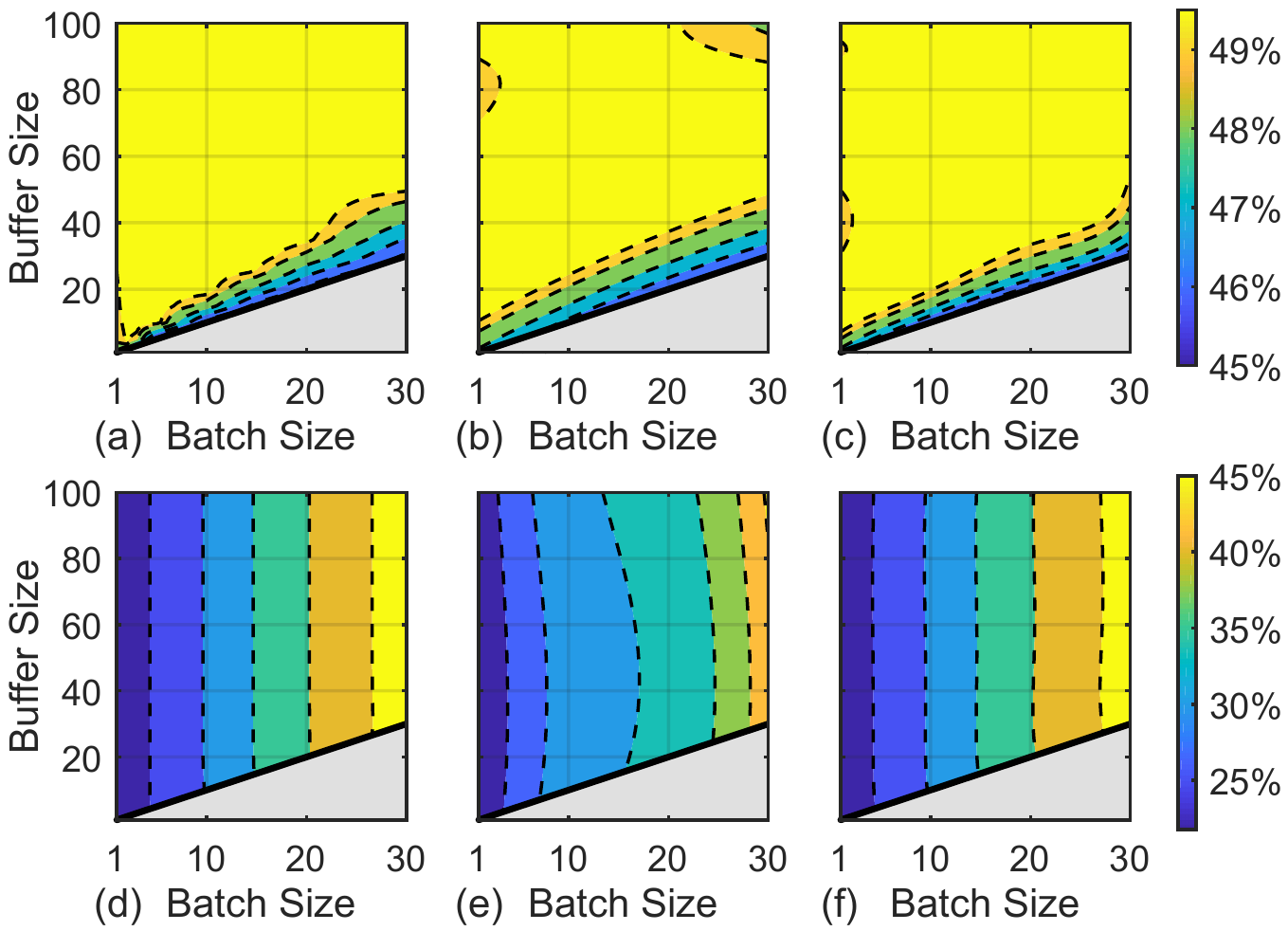}
    \caption{Performance gains (a) actual, and model with degree - (b) 3 and \newline (c) 5. Energy gains (d) actual, and model with degree - (e) 3 and (f) 5.}
    \label{fig:modelContour}
\end{figure}

At testing time, for every memory trace of last 1000 accesses at time $t_i$, the trace feature vector is coupled with the current (at time $t_i$) buffer size ($W_i$) and batch size ($B_i$) to form a vector of size 10 which is sent to the two polynomial models that output $pg_i$ and $eg_i$. These are then used to calculate the reward ($r_i$) as convex combination $\alpha \cdot pg_i + \beta \cdot eg_i$ where the weights $\alpha$ and $\beta$ ($\alpha + \beta = 1$) can be changed based on user requirements targeting either optimal performance or energy (experiments in this work use $\alpha$ = 0.1 and $\beta$ = 0.9 to prioritize energy reduction). This whole pipeline is run multiple times by changing buffer and batch sizes to achieve maximum reward using gradient ascent. For the current model, we used learning rate as 0.01 with 0.9 momentum. This gives the ($W_{i+1}$,$B_{i+1}$) tuple which maximizes the reward (keeping area minimum) and is used for the next interval till $t_{i+1}$. Figure \ref{fig:modelContour} shows actual performance and energy gains with polynomial models of degree 3 and 5. With higher degree, outputs are closer to actual values but require more model parameters (AVAC uses degree 5 models). Finally, the Wait Buffer and batch sizes were fixed to 80 and 10 respectively for the Static RRAM-VAC \cite{Tuli_aspdac}.

\section{Experimental Setup}
\label{sec:ExpSetup}

This section discusses the basic setup, energy and speed assumptions, with biomedical applications and generic benchmarks. The simulations were performed on MATLAB.

\subsection{Energy and Speed Assumptions}

For a programming voltage of 1V to 1.5V, programming time of a few tens of nano-seconds can be achieved \cite{vianello}. Hence, we have assumed a 50 ns worst-case programming time at a programming voltage of 1V. 100 $\mu$A programming current has been assumed for a high ratio of resistance between the High Resistance State (HRS) and the Low Resistance State (LRS) and for several years of memory lifetime \cite{vianello}. We take a normal distribution of programming time with the mean ($\mu$) at 25 ns and a variance ($\sigma$) of 5 ns \cite{Tuli_aspdac}. The AVAC model can be further extended to account for ageing and corresponding variation in the distribution. As explained in \cite{Tuli_aspdac}, energy characterization is done for every bit as an integral ($V \int{I \cdot dt}$) taking into account the shift in the current ($I$) for every stochastic transition (HRS to LRS or vice-versa). WT circuit switching detection time is taken to be 1 ns \cite{Tuli_aspdac}. The read energy is considered as 1 pJ per bit \cite{jain2019}. Scaling down CMOS technology has led to increase in leakage power, primary concern in low-power circuits. The leakage current (for 1V programming voltage) can go as high as 15nA per bit cell (and even higher in some cases) in the current deep-submicron nodes \cite{CAM_leakage2}. This leads to a leakage gain of 480nW per word-line of 32 bits (switching them off dynamically when not needed) in the proposed approach. Implementation of AVAC models described in Sections \ref{sec:fe} and \ref{sec:ml} on Artix-7 FPGA shows that it has 0.278\% energy overhead compared to a 4GB RRAM, which is negligible \cite{RRAM_power}.

\subsection{Applications and Benchmarks}

Experiments were performed on different applications relevant to Edge devices in the biomedical domain. These include Compressed Sensing (CS) for an Electro-Cardiogram (ECG) signal, Feature Extraction (FE) and Decision Tree (DT) in Epilepsy seizure detection algorithm, with Matrix Multiplication (MM) and Convolution (Conv) as two kernels \cite{vasudevan17, Krizhevsky12} as used in \cite{Tuli_aspdac}. DT\_C is the DT application post-processed with an L1 cache of 16 words. We also use Sysbench CPU \cite{kopytov2004sysbench} and Apache \cite{apachebench} to further increase the diversity of applications. All benchmarks traces were extracted on Raspberry-Pi 4 and used to fit the polynomial models PG and EG.

\section{Experimental results}
\label{sec:ExpResults}


\begin{figure}
    \centering
    \includegraphics[width = \columnwidth]{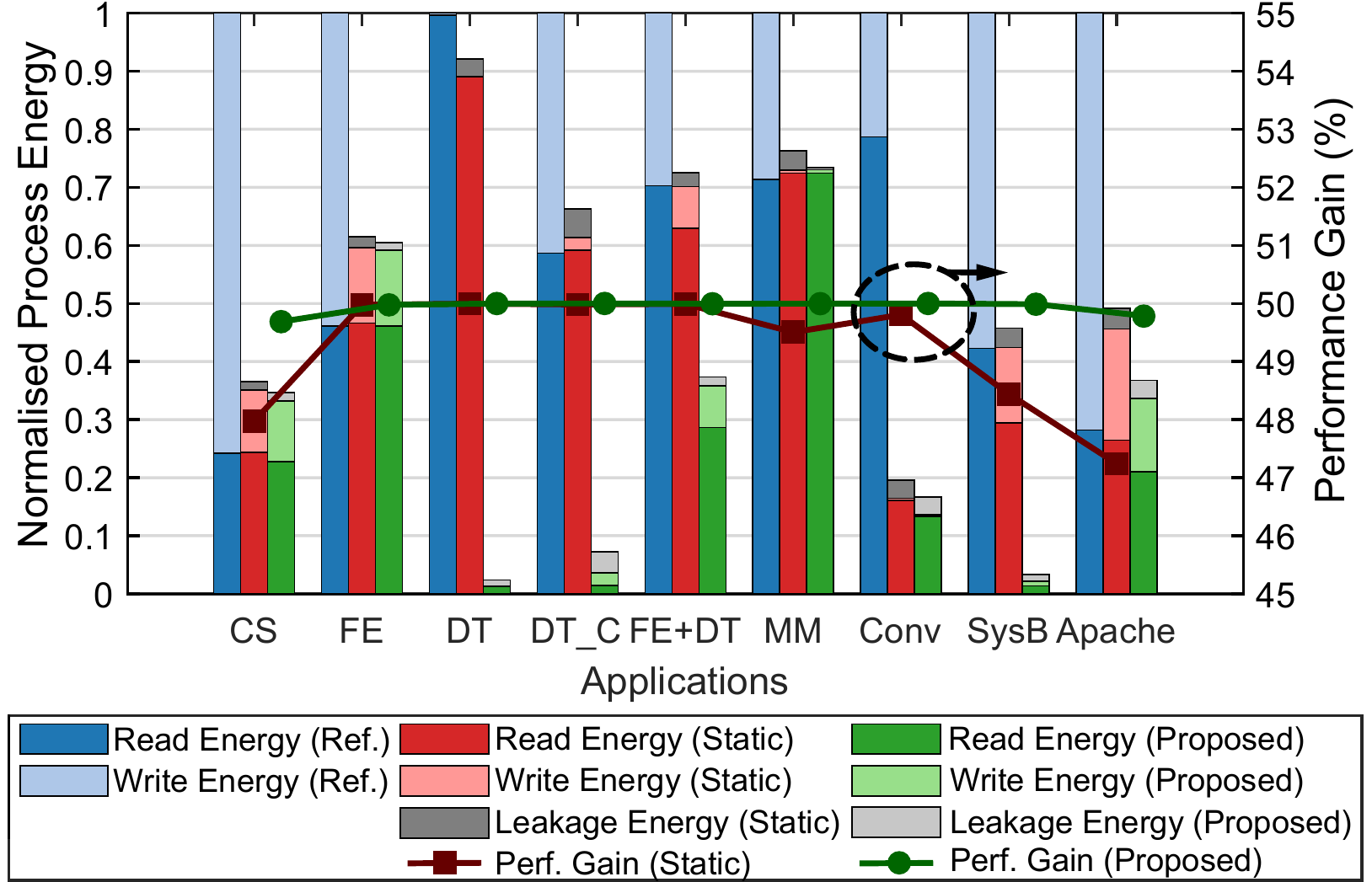}
    \caption{Energy and Performance comparison for different applications}
    \label{fig:energyGain}
\end{figure}

Figure \ref{fig:energyGain} compares the energy and performance gains for the Static and the Adaptive RRAM-VAC with the reference case not using the RRAM-VAC. These gains can be explained by comparing the model parameter optima, i.e the optimal values of the tuple ($W$, $B$) for different applications, with that of the Static RRAM-VAC - (80, 10). For the applications CS and FE, the parameter optima ((80, 60) and (70, 50) respectively) were quite close to that of the Static RRAM-VAC, giving only about 2\% more gains in energy compared to the Static case. The CS and FE have low locality of addresses, and a low read/write ratio. Since the parameters for Static RRAM-VAC were decided based on writes to random addresses \cite{Tuli_aspdac} (implying a low locality and zero read/write ratio) these memory traces are quite close to the assumed one in the Static RRAM-VAC. Still, CS gains a little in performance owing to a larger batch size (averaging the variability in write times). However, MM and Conv show little dependence of the parameters on the gains, with their respective optima at (10, 10) and (80, 80). This is due to the fact that both have balanced read/write ratios. In MM, AVAC reduces leakage power by reducing the Wait Buffer size as locality is low, since a larger Wait Buffer does not provide significant advantage. On the other hand, Conv has high gains both in the Static and Adaptive RRAM-VAC and benefits from a large Wait Buffer to reduce energy of read and write accesses.

AVAC provides high gains in DT, DT\_C and FE+DT, with their respective model optima at - (30,30), (60,60) and (50,50). DT shows maximum improvement in energy gains- 89\% higher than the Static case. With otherwise low locality but high address repetition, these applications perform better with a smaller Wait Buffer. This can be explained by high read/write burst size and high address repetition. Sysbench shows high gains with the model optimum  at (30,30) due to high read/write locality. On the other hand, Apache has similar optimum at (70,50) as the Static case, due to low read/write locality. Both benefit with a larger batch size, providing 3\% higher performance gains compared to the Static case.

\begin{figure}
    \centering
    \includegraphics[width = \columnwidth]{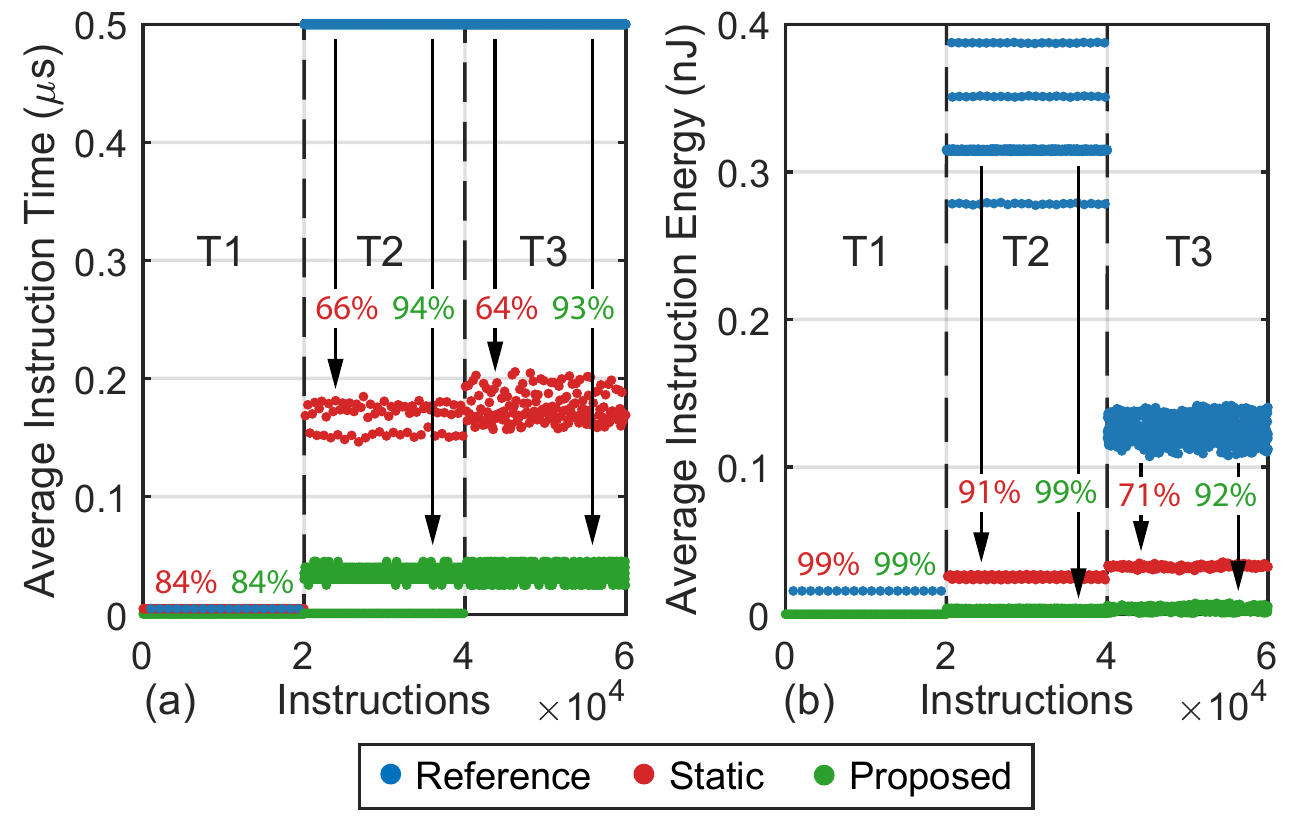}
    \caption{Transient simulations for Static and Adaptive RRAM-VAC for HealthFog application (a) Performance Gains and (b) Energy Gains}
    \label{fig:HealthFog}
\end{figure}

To demonstrate efficacy of AVAC, we further show the performance and energy gains for a real-life Edge application called HealthFog \cite{tuli2019healthfog} which provides high accuracy healthcare services using ensemble deep learning. Figure \ref{fig:HealthFog} shows the application running in 3 major stages: (T1) the program performs large number of ECG read operations and shares data in real-time with sensors and actuators; (T2) the program performs task scheduling and migration decisions for minimum service-level-agreement violations \cite{Gill2019}; (T3) the program utilizes ensemble deep learning based methods to evaluate the data and generate results like health analysis and automated prescription generation. Gains in static and adaptive cases are same in T1 due to only-read operations. T2 has large number of convolution operations giving high energy gains in both the Static and Adaptive cases, compared to the reference. Further, T3 has many MM-like operations and also bootstrapping processes similar to those in the DT, leading to higher gains (29\% performance and 19\% energy). Also, Figure \ref{fig:HealthFog} highlights how the proposed ML-based model can adapt to shifts in memory access patters to instantly enhance gains.

\section{Conclusions}
\label{sec:Conc}

In this work, we proposed the Adaptive RRAM Variability-Aware Controller (AVAC), which uses Machine Learning (ML) techniques to dynamically configure the Wait Buffer and the batch sizes. This not only mitigates the device-to-device variability in RRAMs, but also further optimizes performance and energy dependent on the memory access patters. Gains were simulated for different applications in the Edge/Fog environment, and compared to the Static RRAM-VAC with fixed Wait Buffer and batch sizes. With the considered RRAM technology, AVAC provides up to 94\% gains in performance (up to 29\% more than the Static case) and up to 99\% gains in energy (up to 19\% higher than the Static case). The model can dynamically tune operation parameters in complex and varied Edge and Fog applications to maintain the best performance and energy at all times. Other ML models, with dynamic interval sizing can be investigated as part of future work.





\bibliographystyle{IEEEtran}
\bibliography{biblio}
%



\end{document}